\documentclass[10pt,letterpaper,twocolumn]{article} 

\usepackage{ol2}
\usepackage[draft]{hyperref}
\usepackage{amsmath}

\begin{document}

\twocolumn[ 

\title{Non-reciprocal transmission in photonic lattices based on unidirectional coherent perfect absorption}


\author{Stefano Longhi}

\address{Dipartimento di Fisica, Politecnico di Milano and Istituto di Fotonica e Nanotecnologie del Consiglio Nazionale delle Ricerche, Piazza L. da Vinci 32, I-20133 Milano, Italy (stefano.longhi@polimi.it)}

\begin{abstract}
A method for realizing asymmetric (one-way) transmission of discretized light in modulated, linear and purely passive optical lattices is suggested, which exploits the idea of unidirectional coherent perfect absorption. The system consists of a linear photonic lattice of coupled resonators or waveguides, side coupled to a chain of lossy elements, in which light can avoid the occupation of the dissipative sites when propagating in one way, but not in the opposite one. Non-reciprocity requires modulation of the resonator/waveguide parameters, realizing a {\it dissipative} optical Aharonov-Bohm diode with non-reciprocal behavior.
\end{abstract}



 ] 

\noindent 
Unidirectional transport is at the heart of many fundamental problems and applications in science and technology.
In optics, unidirectional transport enables to realize an optical diode, which is generally obtained by time-reversal symmetry breaking using magnetic fields in linear media or by exploiting nonlinear interactions. In the past few years, several efforts have been devoted toward the realization of miniaturized on-chip optical isolators using novel concepts and design methods. 
These include the use of integrated magneto-optics materials \cite{r1,r2}, time-varying optical structures \cite{r3,r4,r5,r6,r7}, 
 and $\mathcal{PT}$-symmetric nonlinear optical media \cite{r8,r9,r10,r11}. Asymmetric propagation has been also shown to occur 
for discretized light in photonic lattices, i.e. arrays of coupled optical resonators or waveguides, where transport is described by an effective discrete Schr\"{o}dinger equation \cite{r11,r12,r13,r14,r15,r16,r17,r18}. In such studies, one-way transmission was realized by the introduction of some nonlinear and asymmetric defects in the lattice \cite{r11,r12,r13,r16,r17,r18} or using magnetic media \cite{r14,r15}.\par
\begin{figure}[htb]
\centerline{\includegraphics[width=8.4cm]{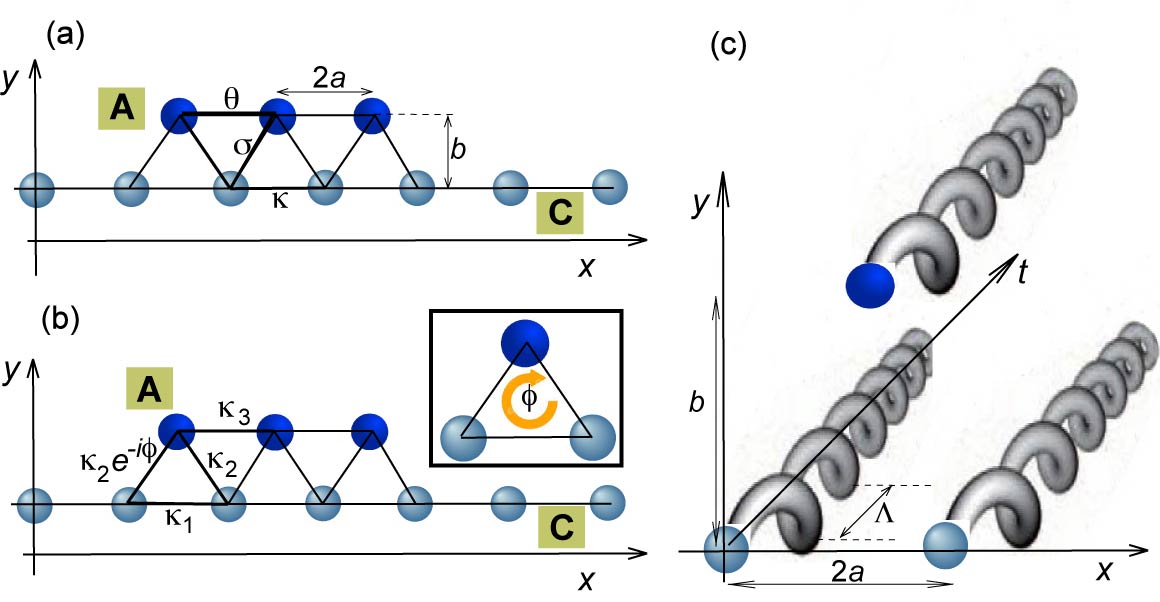}} \caption{ 
\small
(Color online) (a) Schematic of a linear array of coupled resonators or waveguides (C) side-coupled to a chain of $(S+1)$ dissipative sites (A). The resonator frequencies are modulated as discussed in the text. (b) Effective static lattice model obtained in the rotating-wave approximation, with the appearance of a Peierls phase $\phi$, corresponding to an effective magnetic flux $\phi$ threading each triangular plaquette (see the inset). (c) Realization of the Peierls phase using an array of helical waveguides.}
\end{figure}
In this Letter a different route toward the realization of asymmetric transport in discrete optical systems is proposed, which is based on the idea of {\it unidirectional} coherent perfect absorption (CPA). Unidirectional transparency is obtained by realizing a dissipative Aharonov-Bohm diode, in which an effective magnetic field is introduced by periodic modulation of the optical parameters \cite{r6,r19,r20,r21}. The method works for linear and passive lattices and does not require nonlinear, active nor magnetic media.\par 
We consider light propagation in a one-dimensional lattice C, which is side coupled to a number $(S+1)$ of lossy (dissipative) sites A as schematically shown in Fig.1(a). Such a structure can be realized by coupled resonator optical waveguides (CROW) or by arrays of evanescently-coupled optical waveguides. Light transport is described by the coupled-mode equations
 \begingroup
\fontsize{9.3pt}{12pt}\selectfont
\begin{eqnarray}
i\frac{dc_n}{dt}  =   \omega_{n,C} c_n+\kappa (c_{n+1}+c_{n-1})+ \sigma \sum_{l=0}^{S} (\delta_{n,l}+\delta_{n,l+1})a_l  \\
i\frac{da_l}{dt}  =   (\omega_{l,A}-i\gamma) a_l+\sigma (c_{l}+c_{l+1})+ \theta (a_{l+1}+a_{l-1}) 
\end{eqnarray}
\endgroup
($l=0,1,...,S$), with $a_l=0$ for $l <0$ and $l>S$. In Eqs.(1,2), $\omega_{n,C}$ and $\omega_{l,A}$ are the resonator frequencies (propagation constants) of the resonators (waveguides) in the main lattice C and in the side-coupled chain A, respectively, $c_n(t)$ and $a_n(t)$ are the respective field amplitudes, $\kappa$, $\theta$ and $\sigma$ are the coupling constants [as shown in Fig.1(a)], and $\gamma$ is the loss rate in the lossy resonators (waveguides) A. Note that in CROW structures $t$ is the time variable, whereas in waveguide arrays it represents the longitudinal spatial propagation coordinate.  For $S=0$ and for uniform resonator frequencies $\omega_{n,A}=\omega_{A}$, $\omega_{l,B}=\omega_B$, the structure of Fig.1(a) is a typical configuration that displays a Fano resonance in transmittance for $\omega_A \neq \omega_C$. However, for static resonators, i.e. $\omega_{C,A}$ independent of time, the transmittance  (as well as reflectance) is independent of the incidence side. Such a symmetric (reciprocal) behavior holds regardless the number $(S+1)$ of resonators in the chain A nor the value of the loss rate $\gamma$. To obtain a non-reciprocal behavior, we introduce an effective magnetic field for photons by proper modulation of the resonator frequencies $\omega_{n,C}$ and $\omega_{l,A}$ so as to introduce  
Peierls phases in the hopping rates \cite{r6,r19,r20,r21}. Previous works have suggested and discussed in details different methods to realize an effective gauge field for photons in CROW and waveguide lattices \cite{r19,r20,r21,r22,r23,r24}. Here we consider the method of a modulated gradient index ramp, which is inspired by the method of photon-assisted tunneling used to introduce synthetic gauge fields for matter wave systems \cite{r25}. Specifically, let us assume
\begingroup
    \fontsize{9.5pt}{12pt}\selectfont
\begin{equation}
\omega_{n,C}(t)=\omega_C+2nF_x(t) \; , \; \; \omega_{l,A}(t)=\omega_A+(2l+1)F_x(t)+F_y(t) \;\;\;
\end{equation}
\endgroup
where $\omega_{C}$, $\omega_A$ are the static resonator frequencies, 
\begin{equation}
F_x(t)= -\Gamma_x \omega \sin(\omega t+\varphi) \; , \; \; F_y(t)=\Gamma_y \omega \cos(\omega t+\varphi)
\end{equation}
are the sinusoidally-modulated gradient index in the horizontal ($x$) and vertical ($y$) directions, $\omega$ is the modulation frequency, and 
\begin{equation}
\Gamma_x=\Gamma \sin (\varphi) \;, \;\;\; \Gamma_y=\Gamma \cos (\varphi)
\end{equation}
are the modulation amplitudes. In the high-frequency modulation limit $\omega  \gg \kappa, \sigma, \theta$ and assuming the resonance condition $M \omega=\omega_{C}-\omega_A$ for some integer $M$,
 \begin{figure}[htb]
\centerline{\includegraphics[width=8.8cm]{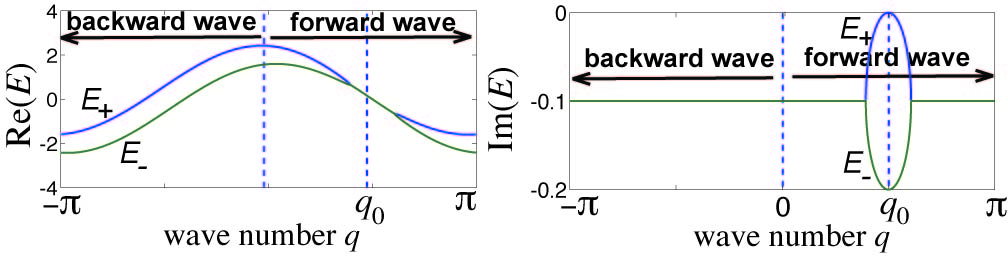}} \caption{ \small
(Color online) Dispersion curves $E_{\pm}(q)$ (real and imaginary parts) of the two minibands for the binary lattice of Fig.1(b) in the $S \rightarrow \infty$ limit. Parameter values are: $\kappa_1=\kappa_3=1$, $\kappa_2=0.3$, $\phi=\pi/2$, and $\gamma=0.2$.}
\end{figure}
 \begin{figure}[htb]
\centerline{\includegraphics[width=7.8cm]{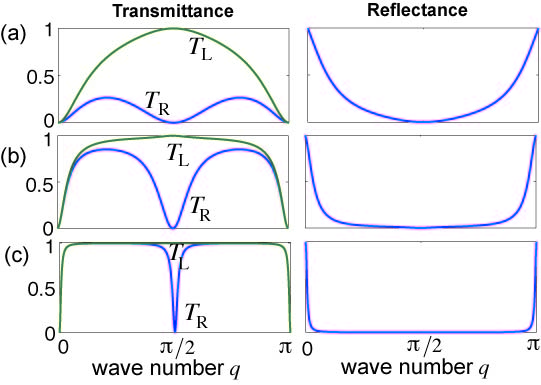}} \caption{ 
\small
(Color online) Spectral transmittance $T_{L,R}(q)=|t_{L,R}(q)|^2$ and reflectance $R(q)=|r_{L}(q)|^2=|r_R(q)|^2$ for left (L) and right (R) incidence sides in the lattice of Fig.1(b) with $S=0$ and for a few values of $\kappa_2/\kappa_1$: (a) $\kappa_2 / \kappa_1=1$, (b) $\kappa_2 / \kappa_1=0.5$ and (c) $\kappa_2 / \kappa_1=0.2$. In the plots the loss rate $\gamma$ is chosen according to the CPA condition (14).}
\end{figure}
an effective stationary lattice with a fictitious magnetic flux is obtained (see e.g. \cite{r19,r20}). In fact, after setting $c_n(t)=C_n(t) \exp[-2in \Theta_x-i \omega_Ct]$, $a_l(t)=A_l(t) \exp[-(2l+1)i\Theta_x-i\Theta_y-i \omega_At]$, where $\Theta_x(t)=\Gamma_x \cos(\omega t+ \varphi)$ and $\Theta_y(t)=\Gamma_y \sin (\omega t+ \varphi)$, under the rotating-wave approximation Eqs.(1) and (2) reduce to the following coupled-mode equations for the amplitudes $C_n$ and $A_l$
\begingroup
    \fontsize{8.5pt}{12pt}\selectfont
 \begin{eqnarray}
i \frac{dC_n}{dt}= \kappa_1 (C_{n+1}+C_{n-1})+\kappa_2 \sum_{l=0}^{S} (\delta_{n,l} \exp(-i \phi)+\delta_{n,l+1}) A_l  \\
i \frac{dA_l}{dt}=-i \gamma A_l+\kappa_3 (A_{l+1}+A_{l-1})+\kappa_2 \exp(i \phi) C_{l}+\kappa_2 C_{l+1} 
\end{eqnarray}
\endgroup
($l=0,1,2,..,S$), where we have set
\begin{eqnarray}
\kappa_1 & = & \kappa \mathcal{J}_0(2 \Gamma \sin \varphi) \;\;, \;\;\;  \kappa_2 = \sigma \mathcal{J}_M(\Gamma) \\
\kappa_3 & = & \theta \mathcal{J}_0(2 \Gamma \sin \varphi) \;\;, \;\;\;
\phi  =   2 M \varphi
\end{eqnarray}
and where $\mathcal{J}_n$ is the Bessel function of the first kind and of order $n$. Equations (6) and (7) describe the  effective stationary lattice shown in Fig.1(b), in which the Peierls phase $\phi$ mimics an effective magnetic flux threading each closed triangular loop. Such an effective magnetic flux, together with the dissipation in the lattice sites A, can provide asymmetric transmission along the main lattice C. To this aim, let us first consider the limiting case $S \rightarrow \infty$, i.e. of two infinitely-extended side-coupled arrays  C and A forming a binary lattice. The band structure $E=E(q)$ of the binary array can be readily obtained from Eqs.(6) and (7) with the Ansatz $C_n(t)=C \exp[-iE(q)t-iqn]$, $A_n(t)=A \exp[-iE(q)t-iqn]$,
where $-\pi \leq q < \pi$ is the Bloch wave number. 
The dispersion relations for the two lattice minibands read
\begingroup
    \fontsize{9pt}{12pt}\selectfont
\begin{eqnarray}
E_{\pm}(q) & = & (\kappa_1+\kappa_3) \cos q -i \frac{\gamma}{2} \\
& \pm & \sqrt{\left[ (\kappa_1-\kappa_3) \cos q + i \frac{\gamma}{2} \right]^2+2 \kappa_2^2 [1+ \cos (q + \phi) ] }. \nonumber
\end{eqnarray}
\endgroup
 \begin{figure}[htb]
\centerline{\includegraphics[width=8.5cm]{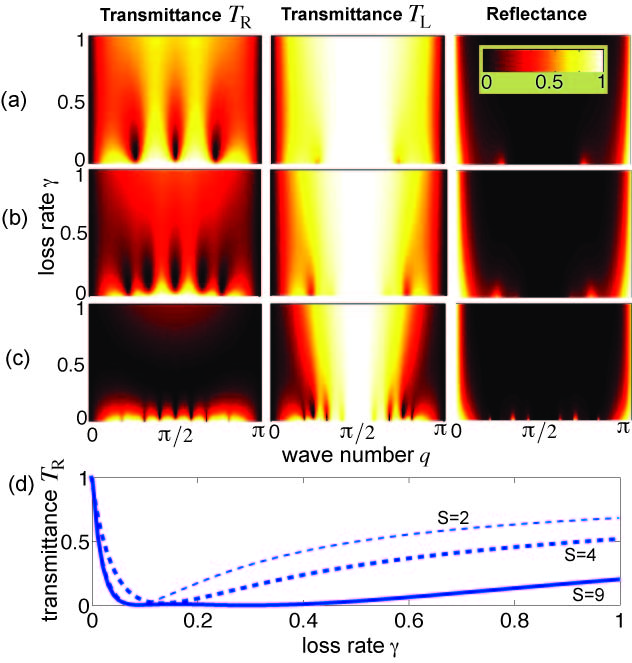}} \caption{ 
\small
(Color online) (a-c) Behavior of the transmittance $T_{L,R}$ (for left and right incidence sides) and reflectance $R$ versus the loss rate $\gamma$ for increasing values of the number $(S+1)$ of lossy sites in the chain C: (a) $S=2$, (b) $S=4$, and (c) $S=9$. Parameter values are: $\kappa_1=\kappa_3=1$, $\kappa_2=0.2$, and $\phi=\pi/2$. In (d) the behavior of the transmittance $T_R(q)$ versus $\gamma$ is shown at $q=\pi/2$.}
\end{figure}
 \begin{figure}[htb]
\centerline{\includegraphics[width=8cm]{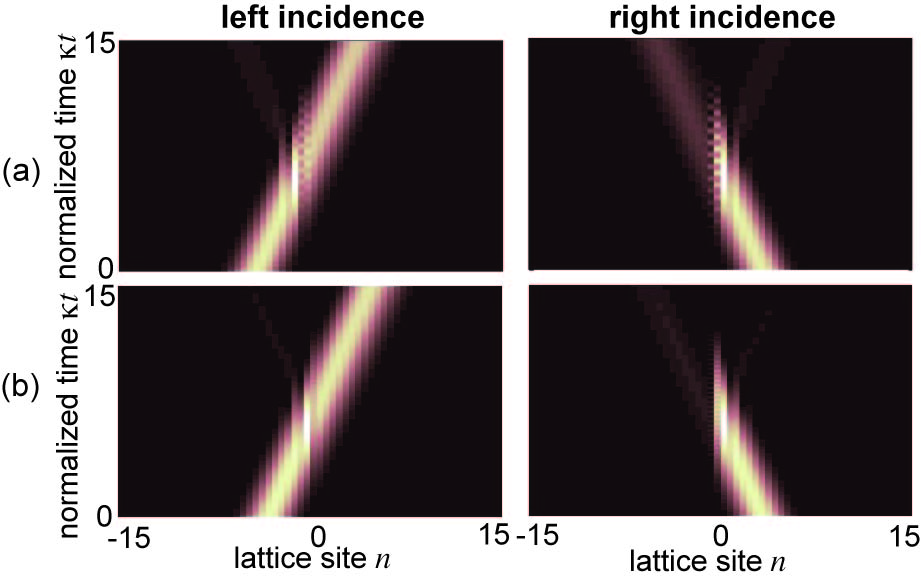}} \caption{ \small
(Color online) Propagation of a Gaussian wave packet of carrier wave number $q= \pi/2$ (snapshot of $|c_n(t)|^2$ in a pseudocolor map), for left and right incidence sides, in the lattice C of Fig.1(a) with modulated parameters and with $S=0$, for (a) $ \omega / \kappa=8$, and (b) $ \omega / \kappa =20$. The other parameter values are given in the text. In (a) the estimated transmittances are $T_L \simeq 89.4 \%$, $T_R \simeq 16 \%$, whereas in (b) one has  $T_L \simeq 95\%$, $T_R \simeq 4.8 \%$.}
\end{figure}
A typical behavior of the real and imaginary parts of $E_{\pm}(q)$ are shown in Fig.2. Note that, for a small coupling $\kappa_2$ between the two sublattices C and A, the miniband $E_+(q)$  corresponds to light localized mainly in the lattice C, whereas the miniband $E_-(q)$ corresponds to light being mostly localized in the (dissipative) lattice A. For $\phi \neq 0$ the symmetry of the dispersion relations under the change $q \rightarrow -q$ is broken, which means that Bloch waves propagating in the forward ($0<q<\pi$) and backward ($-\pi < q <0$) directions in the lattice behave differently. Interestingly, at $q=q_0 \equiv \pi-\phi>0$ the forward propagating wave for the upper miniband ($E_{+}$) is not attenuated, i.e. ${\rm Im}(E_+(q_0))=0$, whereas the backward propagating wave  at $q=-q_0$ it is, i.e. {\it unidirectional absorption} is realized. Such a result indicates that the system behaves non-reciprocally for forward and backward wave propagation, and can be thus exploited to realize an optical diode.\par Let us now consider the case of a finite number $(S+1)$ of dissipative resonators in the chain A. Without loss of generality, let us assume $\kappa_1>0$. The transmission ($t$) and reflection ($r$) coefficients for left (L) and right (R) incidence sides in the array C can be calculated by standard methods. For left-side incidence, 
the scattering solution is of the form $C_n(t)=\exp(-iqn-iEt)+r_L(q,\phi) \exp(iqn-iEt)$ for $n \leq 0$ and $C_n(t)=t_L(q,\phi) \exp[-iq(n-S-1)-iEt] \exp(iqn)$ for $n \geq S+1$, whereas for right-side incidence one has $C_n(t)=t_R(q,\phi) \exp(iqn-iEt)$ for $n \leq 0$ and $C_n(t)= \exp[iq(n-S-1)-iEt]+ r_R(q,\phi) \exp[-iq(n-S-1)-iEt]$ for $n \geq S+1$. In such previous relations, we have set $E=2 \kappa_1 \cos q$ and assumed $0<q<\pi$. One can show that the following general relations hold
\begin{equation}
r_R(q,\phi)=r_L(q,\phi) \; , \;\; t_R(q,\phi)=t_L(q, -\phi).
\end{equation}
 This means that, while the system behaves symmetrically in reflection, it behaves asymmetrically in transmission. A simple analytical form of the reflection/transmission coefficients can be given in case of a single dissipative site. For  $S=0$ one has:
\begingroup
    \fontsize{8pt}{12pt}\selectfont
\begin{eqnarray}
t_L(q,\phi) =  \frac{2 i \sin q \left[ \kappa_1(E+i \gamma) + \kappa_2^2 \exp(i \phi) \right] }{\kappa_1(E+i \gamma)[\exp(2i q)-1] -2 \kappa_2^2 \cos \phi-2 \kappa_2^2 \exp(iq )} \\
r_L(q,\phi) = \frac{2 \kappa_2^2 (\cos q + \cos \phi)}{\kappa_1(E+i \gamma)[\exp(2i q)-1] -2 \kappa_2^2 \cos \phi-2 \kappa_2^2 \exp(iq )}
\end{eqnarray}
\endgroup
where $E=2 \kappa_1 \cos q$. In the following we will consider the case $\phi= \pi/2$, which is the most interesting one, and will omit the dependence of the reflection/transmission coefficients on $\phi$. From Eqs.(12) and (13) one has  $|t_L(q=\pi/2)|=1$, $|t_R(q=\pi/2)|= |\gamma \kappa_1-\kappa_2^2|/(\gamma \kappa_1+\kappa_2^2)$ and $|r_{L,R}(q=\pi/2)|=0$. In particular, provided that the loss rate $\gamma$ satisfies the condition
\begin{equation}
\gamma= \kappa_2^2 / \kappa_1\; ,
\end{equation}
{\it unidirectional} CPA is obtained: namely, a light wave propagating in the forward direction is fully transmitted, without being reflected nor absorbed, whereas a light wave traveling in the backward direction is neither reflected nor transmitted, rather it is fully absorbed in the site A. Hence, when the condition (14) is met, the optical system realizes perfect optical isolation at $q=\pi/2$ because of unidirectional CPA. We not that, while CPA in linear dissipative optical structures generally shows a reciprocal (symmetric) behavior \cite{r26,r27,r28}, i.e. it is independent of the incidence side, in our lattice system CPA is a unidirectional phenomenon because of the presence of the Peierls phase $\phi$. The structure of Fig.1(b) with $S=0$ basically realizes a discrete and {\it dissipative} Aharonov-Bohm interferometer for light waves, allowing for the nonreciprocal transmission of the structure. A typical behavior of the transmittance $T_{R,L}(q)=|t_{L,R}(q)|^2$ and reflectance $R(q)=|r_{R,L}(q)|^2$ for $S=0$ and for a few values of $\kappa_2 / \kappa_1$, corresponding to the CPA condition (14),  is shown in Fig.3. Note that, as $\kappa_2 / \kappa_1$ decreases, the bandwidth around $q= \pi/2$ of unidirectional CPA shrinks. Hence, to realize a relatively broadband optical isolation, a strong coupling of defect A with the main lattice C is required. \par For an arbitrary number  $(S+1)$ of dissipative sites in the chain  A, the condition for unidirectional CPA deviates from Eq.(14), and one has to resort to a numerical analysis to determine the optimum value of $\gamma$. However, as $S$ increases, CPA is observed for a wide range of $\gamma$ values. As an example, Fig.4 shows the behavior of transmittance and reflectance versus the loss rate $\gamma$ for a few values of $S$. Interestingly, even for a small value of $\kappa_2/ \kappa_1$, broadband unidirectional CPA can be realized increasing the value of $S$ [see Fig.4(c)]. \par 
Non-reciprocal transmission predicted by the effective lattice model of Fig.1(b), i.e. in the rotating-wave approximation, has been confirmed by beam propagation simulations within the original modulated lattice model [Eqs.(1) and (2)]. As an example, Fig.5 shows the numerically-computed propagation of a Gaussian wave packet, for either left or right incidence sides, with carrier wave number $q=\pi/2$ and for parameter values $S=0$, $\sigma / \kappa =1.6$, $M=1$, $\varphi=\pi/4$ (corresponding to the magnetic flux $\phi=\pi/2$), $\Gamma=1$, $\omega / \kappa=8$ in Fig.5(a), and $\omega / \kappa=20$ in Fig.5(b). The loss rate $\gamma / \kappa=0.88$ has been tuned to satisfy the CPA condition (14), where the values of $\kappa_{1} / \kappa \simeq 0.56$ and $\kappa_2 / \kappa \simeq 0.7$ are estimated using Eq.(8). The simulations clearly show strong asymmetric transmission, even for a relatively low modulation frequency [Fig.5(a)]. \par
Finally, let us briefly discuss a possible experimental realization of asymmetric transmission exploiting unidirectional CPA and based on spatial light transport in lattices of  dielectric optical waveguides \cite{r29}. In such an optical system, the Peierls phase $\phi$ can be  implemented by periodically rotating the optical axis of waveguides along the propagation direction \cite{r22,r24,r29}; see Fig.1(c). Taking an helical path of the waveguide axis with spatial frequency $\omega= 2 \pi / \Lambda$ described by the parametric equations $x_0(t)= X_0 \sin (\omega t + \varphi)$ and $y_0(t)= -Y_0 \cos (\omega t + \varphi)$, 
 the effective index gradients given in Eqs.(3) and (4) are realized with $\Gamma_x=2 \pi n_s a X_0 \omega / \lambda$ and $\Gamma_y=2 \pi n_s b Y_0 \omega / \lambda$, where $\lambda$ is the optical wavelength (in vacuum) of propagating waves,  $a$ and $b$ are the horizontal and vertical displacements of waveguides in the lattice [see Fig.1(c)], and $n_s$ is the substrate (cladding) refractive index \cite{r29,r30}. For example, let us consider propagation of visible light ($\lambda=633$ nm)  in dielectric helical waveguides manufactured in fused silica by femtosecond laser writing \cite{r24,r30}. Assuming $n_s \simeq 1.46$, $2a=16 \; \mu$m, $b=4 \; \mu$m, $\kappa \sim 1.5 \; {\rm cm}^{-1}$, the simulation of Fig.5(a) corresponds to a helix with period $\Lambda \simeq 0.52$ cm and amplitudes $X_0 \simeq 5 \; \mu$m, $Y_0 \simeq 10 \; \mu$m, a loss rate $\gamma \simeq 1.33 \; {\rm cm}^{-1}$ (i.e. $\simeq 11.5$ dB/cm), and a full propagation length $ t \simeq 10$ cm. The detuning $\omega_C-\omega_A= \omega$ of the modal propagation constant of waveguide A from the main array C corresponds to a refractive index change $\delta n \simeq \lambda/ \Lambda \simeq 1.21 \times 10^{-4}$ of the core of waveguide A, which can be obtained by control of the laser writing speed in the sample. Finally, the required loss rate in waveguide A can be realized, for example, by proper waveguide segmentation. \par
 In conclusion, a method for realizing asymmetric (one-way) transmission of discretized light in modulated optical lattices has been suggested, which is based on the idea of unidirectional coherent perfect absorption. The method works for linear and passive lattices and does not require nonlinear, active nor magnetic media.

\newpage

\footnotesize {\bf References with full titles}\\
\\
\noindent
1. M.A. Levy, {\it Nanomagnetic route to bias-magnet-free, on-chip Faraday rotators}
J. Opt. Soc. Am. B {\bf 22}, 254 (2005).\\
2.  L. Bi, J. Hu, P. Jiang, D.H. Kim, G.F. Dionne,	 L.C. Kimerling, and C.A. Ross, {\it On-chip optical isolation in monolithically integrated non-reciprocal
optical resonators}, Nat. Photonics {\bf 5}, 758 (2011).\\
3. Z. Yu and S. Fan, {\it Complete optical isolation created by indirect interband photonic transitions}, Nat. Photonics {\bf 3}, 91 (2009).\\
4. M.S.  Kang, A. Butsch, and P. St. J. Russell, {\it Reconfigurable light-driven opto-acoustic
isolators in photonic crystal fibre}, Nat. Photonics {\bf 5}, 549 (2011).\\
5. H. Lira, Z. Yu, S. Fan, and M. Lipson, {\it Electrically Driven Nonreciprocity Induced by Interband Photonic Transition on a Silicon Chip},
Phys. Rev. Lett. {\bf 109}, 033901 (2012).\\
6. K. Fang, Z. Yu, and S. Fan, {\it Photonic Aharonov-Bohm Effect Based on Dynamic Modulation}, Phys. Rev. Lett. {\bf 18}, 153901 (2012).\\
7. L.D. Tzuang, K. Fang, P. Nussenzveig, S. Fan, and M. Lipson, {\it Non-reciprocal phase shift induced by an effective magnetic
flux for light}, Nat. Photonics {\bf 8}, 701 (2014).\\
8. N. Bender, S. Factor, J. D. Bodyfelt, H. Ramezani, D. N. Christodoulides, F. M. Ellis, and T. Kottos, {\it Observation of Asymmetric Transport in Structures with Active Nonlinearities}, Phys. Rev. Lett. {\bf 110}, 234101 (2013).\\
9. L. Chang,	 X. Jiang, S. Hua, C. Yang, J. Wen, L. Jiang, G. Li, G. Wang, and M. Xiao, {\it Parity-time symmetry and variable optical isolation in active-passive-coupled microresonators}, Nat. Photonics {\bf 8}, 524 (2014).\\
10. B. Peng, S. K.Ozdemir, F. Lei, F. Monifi, M. Gianfreda, G.L. Long, S. Fan, F. Nori, C.M. Bender, and L. Yang,
{\it Parity-time-symmetric whispering-gallery microcavities}, Nat. Phys. {\bf 10}, 1 (2014).\\
11. F. Nazari, N. Bender, H. Ramezani, M.K. Moravvej-Farshi, D. N. Christodoulides, and T. Kottos, {\it Optical isolation via $\mathcal{PT}$-symmetric nonlinear Fano resonances}  
Opt. Express {\bf 22}, 9574 (2014).\\
12. S. Lepri and G. Casati, {\it AsymmetricWave Propagation in Nonlinear Systems}, Phys. Rev. Lett. {\bf 106}, 164101 (2011).\\
13. S. Lepri and B. Malomed, {\it  Symmetry breaking and restoring wave transmission in diode-antidiode double chains}, Phys. Rev. E {\bf 87}, 042903 (2013).\\
14. P. Kumar and M. Levy, {\it On-chip optical isolation via unidirectional Bloch oscillations in a waveguide array}, Opt. Lett. {\bf 37}, 3762 (2012).\\
15. R. El-Ganainy, A. Eisfeld, M. Levy, and D. N. Christodoulides, {\it On-chip non-reciprocal optical devices based on quantum inspired photonic lattices}, Appl. Phys. Lett. {\bf 103}, 161105 (2013).\\
16. N. Li and J. Ren, {\it Non-Reciprocal Geometric Wave Diode by Engineering Asymmetric Shapes of Nonlinear Materials}, Sci. Rep. {\bf 4}, 6228 (2014).\\
17. Y. Xu and A.E. Miroshnichenko, {\it Reconfigurable nonreciprocity with a nonlinear Fano diode}, Phys. Rev. B {\bf 89}, 134306 (2014).\\
18. T.F. Assuncao, E.M. Nascimento, and M.L. Lyra, {\it Nonreciprocal transmission through a saturable nonlinear asymmetric dimer}, Phys. Rev. E {\bf 90}, 022901 (2014).\\
19. S. Longhi, {\it Effective magnetic fields for photons in waveguide and coupled resonator lattices}, Opt. Lett. {\bf 38}, 3570 (2013).\\
20. S. Longhi, {\it Aharonov-Bohm photonic cages in waveguide and coupled resonator lattices by synthetic magnetic fields}, Opt. Lett. {\bf 39}, 5892 (2014).\\
21. K. Fang, Z. Yu, and S. Fan, {\it Realizing effective magnetic field for photons by controlling the phase of dynamic modulation}, Nature Photon. {\bf 6}, 782 (2012).\\
22. S. Longhi, {\it Bloch dynamics of light waves in helical optical waveguide arrays}, Phys. Rev. B {\bf 76}, 19511 (2007).\\
23. Q. Lin and S. Fan, {\it Light Guiding by Effective Gauge Field for Photons}, Phys. Rev. X {\bf 4}, 031031 (2014).\\
24. M. C. Rechtsman, J.M. Zeuner, Y. Plotnik, Y. Lumer, D. Podolsky, F. Dreisow, S.Nolte, M. Segev, and A. Szameit, {\it Photonic Flouet Topological Insulators},
Nature {\bf 496}, 196 (2013).\\
25. A. Bermudez, T. Schaetz, and D. Porras, {\it Synthetic Gauge Fields for Vibrational Excitations of Trapped ions}, Phys. Rev. Lett. {\bf 107}, 150501 (2011).\\
26.  A. Yariv, {\it Universal relations for coupling of optical power between microresonators and dielectric waveguides}, Electron. Lett. {\bf 36}, 321 (2000).\\
27. M. Cai, O. Painter, and K. J. Vahala, {\it Observation of critical coupling in a fiber taper to a silica-microsphere whispering-gallery mode system}, Phys. Rev. Lett. {\bf 85}, 74 (2000).\\
28. Y.  Chong, L. Ge, Li, H. Cao, and A. Stone, {\it Coherent Perfect Absorbers: Time-Reversed Lasers}, Phys. Rev. Lett. {\bf 105}, 053901 (2010).\\
29.  I.L. Garanovich, S. Longhi, A.A. Sukhorukov, and Y.S. Kivshar, {\it Light propagation and localization in modulated photonic lattices and waveguides},
Phys. Rep. {\bf 518}, 1 (2012).\\
30. A. Crespi, G. Corrielli, G. Della Valle, R. Osellame, and S. Longhi, {\it Dynamic band collapse in photonic graphene}, New J. Phys. {\bf 15}, 013012 (2013).\\


\begin{thebibliography}{99}



\bibitem{r1}
M.A. Levy, J. Opt. Soc. Am. B {\bf 22}, 254 (2005).
\bibitem{r2}
L. Bi, J. Hu, P. Jiang, D.H. Kim, G.F. Dionne, L.C. Kimerling, and C.A. Ross, Nat. Photonics {\bf 5}, 758 (2011).
\bibitem{r3}
Z. Yu and S. Fan, Nat. Photonics {\bf 3}, 91 (2009).
\bibitem{r4}
M.S.  Kang, A. Butsch, and P. St. J. Russell, Nat. Photonics {\bf 5}, 549 (2011).
\bibitem{r5}
H. Lira, Z. Yu, S. Fan, and M. Lipson, Phys. Rev. Lett. {\bf 109}, 033901 (2012).
\bibitem{r6}
K. Fang, Z. Yu, and S. Fan, Phys. Rev. Lett. {\bf 18}, 153901 (2012).
\bibitem{r7}
L.D. Tzuang, K. Fang, P. Nussenzveig, S. Fan, and M. Lipson, Nat. Photonics {\bf 8}, 701 (2014). 
\bibitem{r8}
N. Bender, S. Factor, J. D. Bodyfelt, H. Ramezani, D. N. Christodoulides, F. M. Ellis, and T. Kottos, Phys. Rev. Lett. {\bf 110}, 234101 (2013).
\bibitem{r9}
L. Chang,	 X. Jiang, S. Hua, C. Yang, J. Wen, L. Jiang, G. Li, G. Wang, and M. Xiao, Nat. Photonics {\bf 8}, 524 (2014).
\bibitem{r10}
B. Peng, S. K.Ozdemir, F. Lei, F. Monifi, M. Gianfreda, G.L. Long, S. Fan, F. Nori, C.M. Bender, and L. Yang, Nat. Phys. {\bf 10}, 1 (2014).
\bibitem{r11}
F. Nazari, N. Bender, H. Ramezani, M.K. Moravvej-Farshi, D. N. Christodoulides, and T. Kottos, Opt. Express {\bf 22}, 9574 (2014).
\bibitem{r12}
S. Lepri and G. Casati, Phys. Rev. Lett. {\bf 106}, 164101 (2011).
\bibitem{r13}
S. Lepri and B. Malomed, Phys. Rev. E {\bf 87}, 042903 (2013).
\bibitem{r14}
P. Kumar and M. Levy, Opt. Lett. {\bf 37}, 3762 (2012).
\bibitem{r15}
R. El-Ganainy, A. Eisfeld, M. Levy, and D. N. Christodoulides, Appl. Phys. Lett. {\bf 103}, 161105 (2013).
\bibitem{r16}
N. Li and J. Ren, Sci. Rep. {\bf 4}, 6228 (2014).
\bibitem{r17}
Y. Xu and A.E. Miroshnichenko, Phys. Rev. B {\bf 89}, 134306 (2014).
\bibitem{r18}
T.F. Assuncao, E.M. Nascimento, and M.L. Lyra, Phys. Rev. E {\bf 90}, 022901 (2014).
\bibitem{r19}
S. Longhi, Opt. Lett. {\bf 38}, 3570 (2013).
\bibitem{r20}
S. Longhi, Opt. Lett. {\bf 39}, 5892 (2014).
\bibitem{r21}
K. Fang, Z. Yu, and S. Fan, Nat. Photonics {\bf 6}, 782 (2012).
\bibitem{r22}
S. Longhi, Phys. Rev. B {\bf 76}, 19511 (2007).
\bibitem{r23}
Q. Lin and S. Fan, Phys. Rev. X {\bf 4}, 031031 (2014).
\bibitem{r24}
M. C. Rechtsman, J.M. Zeuner, Y. Plotnik, Y. Lumer, D. Podolsky, F. Dreisow, S.Nolte, M. Segev, and A. Szameit,
Nature {\bf 496}, 196 (2013).
\bibitem{r25}
A. Bermudez, T. Schaetz, and D. Porras, Phys. Rev. Lett. {\bf 107}, 150501 (2011).
\bibitem{r26}
 A. Yariv,  Electron. Lett. {\bf 36}, 321 (2000).
\bibitem{r27} 
 M. Cai, O. Painter, and K. J. Vahala, Phys. Rev. Lett. {\bf 85}, 74 (2000).
 \bibitem{r28}
Y.  Chong, L. Ge, Li, H. Cao, and A. Stone, Phys. Rev. Lett. {\bf 105}, 053901 (2010).
\bibitem{r29}
I.L. Garanovich, S. Longhi, A.A. Sukhorukov, and Y.S. Kivshar, Phys. Rep. {\bf 518}, 1 (2012).
\bibitem{r30}
A. Crespi, G. Corrielli, G. Della Valle, R. Osellame, and S. Longhi, New J. Phys. {\bf 15}, 013012 (2013).



\end{thebibliography}
\end{document}